\documentclass[aps,
prb,
reprint,
preprintnumbers,
superscriptaddress,
footinbib,
amsfonts,
showkeys,
amssymb,
amsmath,
intlimits,
]{revtex4-2}

\usepackage{bm,latexsym,mathrsfs,enumerate,bbm}
\usepackage{upgreek}
\usepackage[table,x11names]{xcolor}
\usepackage[breaklinks=true,
unicode=true,
urlcolor = RoyalBlue4,
colorlinks = true,
citecolor = RoyalBlue4,
linkcolor = RoyalBlue4
]{hyperref}
\usepackage{graphicx}

\usepackage[labelformat=simple]{subcaption}

\usepackage{chemformula}
\usepackage{siunitx}
\usepackage{enumitem}
\usepackage{mathtools}
\usepackage[cal=boondoxo,scr=rsfs]{mathalfa}
\renewcommand{\vec}[1]{\bm{#1}}
\begin{document}
\title{Impact of curvature-induced Dzyaloshinskii--Moriya interaction on magnetic vortex texture in spherical caps}
	
\author{Mykola I. Sloika}
\email{sloika.m@gmail.com}
\affiliation{Taras Shevchenko National University of Kyiv, 01601 Kyiv, Ukraine}
	
\author{\boxed{\text{Yuri Gaididei}}}
\affiliation{Bogolyubov Institute for Theoretical Physics of National Academy of Sciences of Ukraine, 03680 Kyiv, Ukraine}
	
\author{Volodymyr P. Kravchuk}
\email{volodymyr.kravchuk@kit.edu}
\affiliation{Institut f\"ur Theoretische Festk\"orperphysik, Karlsruher Institut f\"ur Technologie, D-76131 Germany}
\affiliation{Bogolyubov Institute for Theoretical Physics of National Academy of Sciences of Ukraine, 03680 Kyiv, Ukraine}
	
\author{Oleksandr~V.~Pylypovskyi}
\email{o.pylypovskyi@hzdr.de}
\affiliation{Helmholtz-Zentrum Dresden-Rossendorf e.V., Institute of Ion Beam Physics and Materials Research, 01328 Dresden, Germany}
\affiliation{Kyiv Academic University, 03142 Kyiv, Ukraine}
	
\author{Denys Makarov}
\email{d.makarov@hzdr.de}
\affiliation{Helmholtz-Zentrum Dresden-Rossendorf e.V., Institute of Ion Beam Physics and Materials Research, 01328 Dresden, Germany}
	
\author{Denis D. Sheka}
\email[correspondent author: ]{sheka@knu.ua}
\affiliation{Taras Shevchenko National University of Kyiv, 01601 Kyiv, Ukraine}
\date{June 14, 2022}
%
%

\begin{abstract}
	Geometric curvature of nanoscale magnetic shells brings about curvature-induced anisotropy and Dzyaloshinskii-Moriya interaction (DMI). Here, we derive equations to describe the profile of the magnetic vortex state in a spherical cap. We demonstrate that the azimuthal component of magnetization acquires a finite tilt at the edge of the cap, which results in the increase of the magnetic surface energy. This is different compared to the case of a closed spherical shell, where symmetry of the texture does not allow any tilt of magnetization at the equator of the sphere. Furthermore, we analyze the size of the vortex core in a spherical cap and show that the presence of the curvature-induced DMI leads to the increase of the core size independent of the product of the circulation and polarity of the vortex. This is in contrast to the case of planar disks with intrinsic DMI, where the preferred direction of circulation as well as the decrease or increase of the size of vortex core is determined by the sign of the product of the circulation and polarity with respect to the sign of the constant of the intrinsic DMI.
\end{abstract}

\keywords{Curvilinear Micromagnetism; Curved Magnets; Magnetic Caps; Magnetic Vortex; Dzyaloshinskii-Moriya interaction}

\pacs{75.75.-c, 75.78.Cd}



\maketitle


\section{Introduction}

Ferromagnetic particles of submicron size (nanomagnets) demonstrate variety of magnetic properties, which make them suitable for wide range of magnonic, spintronic and spinorbitronic applications including memory and logic devices~\cite{Parkin08,Wiesendanger16,Fert17,Sander17,Vedmedenko20,Back20}. Magnetization textures of nanomagnets strongly depend on the shape and size and, for the case of soft magnetic materials, are typically determined by the interplay between the short range exchange and long range magnetostatic interactions. A representative example is magnetic vortex --- a magnetic topological soliton, which is characterized by the in-plane magnetization circulation following the edge of a confined nanomagnet and possessing an out-of-plane magnetization component at the location of the vortex core \cite{Cowburn00b,Guslienko08b,Hubert09,Fernandez17}. Magnetic vortices have been found to be equilibrium or even ground states in ferromagnets of different shape including planar disks~\cite{Cowburn00b}, rings \cite{Klaui03a, Kravchuk07}, spherical caps \cite{Sheka13b, Streubel12a, Mitin14, Nissen15, Streubel16}, spherical shells \cite{Kravchuk12b, Sloika17}, tori~\cite{Carvalho-Santos10, Vojkovic16a} and more complex geometries~\cite{Carvalho-Santos13, Mancilla-Almonacid20}. 

Geometrical curvature emerged as an appealing possibility to tune anisotropic and chiral responses of magnetic thin films~\cite{Makarov22} and wires~\cite{Sheka22}. In particular, curvature brakes the spatial inversion symmetry and leads to such phenomena as topological patterning \cite{Landeros10, Kravchuk16a, Sloika17, Teixeira19, Stano18b, Mancilla-Almonacid20, Castillo-Sepulveda21} and geometrical magnetochiral effects \cite{Hertel13a, Streubel16a, Dietrich08, Yan11a, Yan12, Otalora16}, for review see \cite{Sheka21b}. For the case of magnetic vortices, curvature-induced magnetochirality is reflected in the coupling between the directions of in-surface magnetization curling and out-of-plane magnetization component, circulation and polarity, respectively~\cite{Kravchuk12a,Sloika14}. These non-reciprocal effects are results of the local curvature-induced Dzyaloshinskii--Moriya interaction (DMI)~\cite{Gaididei14,Sheka15} or nonlocal effects stemming from magnetostatics~\cite{Otalora16,Sheka20a}. Similar to the intrinsic DMI~\cite{Dzyaloshinsky57,Moriya60,Fert17}, its curvature-induced counterpart favors canted spin textures like domain walls~\cite{Pylypovskyi15b}, skyrmions~\cite{Kravchuk16a,Kravchuk18a} and skyrmionium states~\cite{Pylypovskyi18a}. It is already known that intrinsic DMI results in the change of the size of the magnetic vortex core as well in the selection of the  preferred circulation of the in-plane magnetization~\cite{Butenko09}. 

Here, we study the influence of the curvature-induced DMI on the magnetic vortex state in spherical caps made of intrinsically achiral and isotropic ferromagnet. Similarly to spherical shells~\cite{Kravchuk12a}, the effective DMI in a spherical cap leads to the additional tilt of the azimuthal angle of magnetization. The presence of edge for the spherical cap allows non-zero tilt angle of magnetization at the edge. Furthermore, we show that the effective DMI in spherical caps increases the size of the magnetic vortex core for any combination of vortex polarity and circulation. This effect is different compared to the case of a planar disk, where the intrinsic DMI can either decrease or increase the vortex core size dependent on the sign of the product of circulation and polarity with respect to the sign of the DMI constant~\cite{Butenko09}.


\section{Model of a spherical cap}
\label{sec:ModelOfSphericalCap}

The first theory of curvilinear magnetism was formulated by \citet{Gaididei14} to describe local effects in curved magnetic shells. A unified description of the curvature–induced effects is based on a micromagnetic framework of curvilinear magnetism \cite{Sheka20}, which allows to treat local and nonlocal interactions on equal footing. In this work, we will discuss magnetic properties of ultrathin ferromagnetic shells of magnetically soft material, taking into account exchange and magnetostatic interactions only: $E = \int \mathrm{d}\vec{r} \left(\mathscr{E}^{\text{x}} + \mathscr{E}^{\text{ms}}\right)$. The energy density of the isotropic exchange interaction is written as $\mathscr{E}^{\text{x}} = -A \vec{m}\cdot \vec{\nabla}^2\vec{m}$ with $\vec{m}$ being the normalized magnetization and $A$ being the exchange constant. It is established that not only for the case of ultrathin planar films \cite{Gioia97,Carbou01,Kohn05a} but also for ultrathin geometrically curved shells the magnetostatic energy, $ \mathscr{E}^{\text{ms}}$, can be reduced to the shape-induced easy-surface anisotropy $\mathscr{E}_{\textsc{a}}^{\text{ms}}$ \cite{Slastikov05,Fratta16}. In our model, the sample geometry enters the description only through the anisotropy term, $\mathscr{E}_{\textsc{a}}^{\text{ms}} = K \left(\vec{m}\cdot \vec{\hat{n}}\right)^2$, where the effective anisotropy constant $K=2\pi M_{\textsc{s}}^2$ with $M_{\textsc{s}}$ being the saturation magnetization. Here, a coordinate-dependent unit normal vector $\vec{\hat{n}} = \vec{\hat{n}}(\vec{r})$ determines the direction of the magnetic hard axis.

The current micromagnetic approach requires that the sample has a constant thickness $h$ along the normal. This make it possible to suppose that the magnetization does not depend on the normal coordinate. This assumption is valid in the limit of ultrathin shells, namely the shell thickness $h$ should be much smaller than the typical magnetic length scale $\ell = \sqrt{A/K}$. One more restriction is that $h$ should be much smaller than the typical curvature radii $L$. Then, one can systematize geometrical effects by restructuring
all magnetic energy terms according to their local spatial symmetry
\begin{equation} \label{eq:energy-curved}
E = A h \!\! \int_S \!\! \mathrm{d}S \left(\mathscr{E}_0 + \mathscr{E}_{\textsc{a}} + \mathscr{E}_{\textsc{d}}\right).
\end{equation}
The first energy density contribution $\mathscr{E}_0=\eth_\alpha m_i \eth_\alpha m_i$ is a regular isotropic part of the exchange energy, it is similar to the one in a planar film. Here $m_i$ is an $i$th
magnetization component in the curvilinear orthonormal Darboux three-frame $\{\vec{e}_1, \vec{e}_2, \vec{\hat{n}}\}$ on the surface, where $\vec{e}_1$ and $\vec{e}_2$ are unit vectors corresponding to the principal directions. Here and below we use Greek letters $\alpha, \beta, \gamma$ run values 1, 2 and refer to the curvilinear coordinates on the surface; to indicate all three components of any vector, we use Latin indices $i, j = 1, 2, n$. The Einstein summation convention is also assumed. The notation $\eth_\alpha$ is used for tangential derivatives, $\eth_\alpha m_\beta = \left( \mathfrak{g}_{\alpha \alpha}\right)^{-1/2} \left[\partial_\alpha m_\beta + \epsilon_{\beta\gamma} \left(\vec{e}_1 \cdot \partial_\alpha \vec{e}_2\right) m_\gamma \right]$ and $\eth_\alpha m_n = \left( \mathfrak{g}_{\alpha \alpha}\right)^{-1/2} \partial_\alpha m_n$ with $\mathfrak{g}_{\alpha\beta}$ being the metric tensor and $\epsilon_{\beta\gamma}$ being totally antisymmetric tensor \cite{Sheka20a}.

The second term, an effective anisotropy, $\mathscr{E}_{\textsc{a}}$ comprises the shape-induced anisotropy $\mathscr{E}_{\textsc{a}}^{\text{ms}}$ and the curvature-induced  exchange-driven biaxial anisotropy $K_{ij}m_im_j$, where $K_{ij}$ have a bilinear form with respect to the principle curvatures $\kappa_1$ and $\kappa_2$ of the surface, $K_{ij}=A\kappa_i^2\delta_{ij}$ with $\kappa_n^2\equiv \kappa_1^2+\kappa_2^2$. The curvature-induced  exchange-driven DMI $\mathscr{E}_{\textsc{d}} = D_{\alpha}  \mathcal{L}_{\alpha n}^{(\alpha)}$ is determined by the curvilinear analogue of Lifshitz invariants $\mathscr{L}_{ij}^{(\alpha)} = m_i \eth_\alpha m_j - m_j \eth_\alpha m_i$ \cite{Sheka20a}. It is worth noting that in the general case of arbitrary surface, the intensity of the curvature-induced DMI depends on two principal curvatures, $D_{\alpha}=2A \kappa_\alpha$.

\begin{figure}
	\includegraphics[width=0.8\columnwidth]{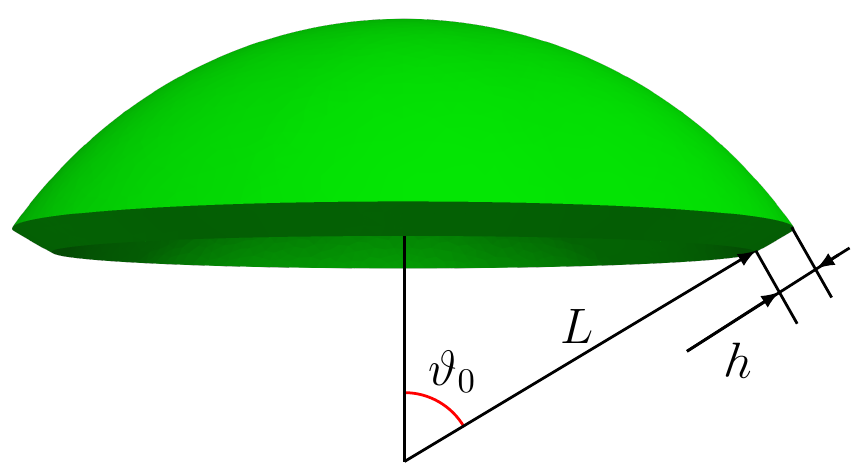}
	\caption{(Color online) \textbf{Schematic of a spherical cap.} The cap is formed by two spherical surfaces with the inner sphere having radius $L$. The thickness of the cap is $h$. The edge surface of the cap is formed by the intersection of the two spherical surfaces by a cone with the opening angle $\vartheta_0$.}
	\label{fig:SphericalCap}
\end{figure}

In the current study, we limit our consideration to a spherical cap structure with the inner radius $L$, the cut angle $\vartheta_0$ and thickness $h\ll\ell\ll L$, see Fig.~\ref{fig:SphericalCap}. In the case of a spherical surface, both principal curvatures read $\kappa_{1,2} = 1/L$. Then the effective anisotropy $\mathscr{E}_{\textsc{a}}=K_{\text{ef}} \left(\vec{m}\cdot \vec{\hat{n}}\right)^2$ with the effective anisotropy constant $K_{\text{ef}}=A/\ell^2+A/L^2$. Both curvature-induced DMI coefficients are equal and inversely proportional to the curvature radius, $D_\alpha=2A/L$.

The ground state of a spherical cap is known to depend on the material and geometrical parameters of the sample \cite{Sheka13b, Streubel12}. In particular, while the vortex state is observed for caps of large diameters, onion state is realized for thin caps of small diameter and uniform easy--axis state is favourable for caps of small diameter and large thickness.


\section{In-surface magnetization curling}
\label{sec:curling}

It was predicted by \citet{Butenko09, Butenko10a} and experimentally confirmed by \citet{Im12}, \textit{intrinsic} DMI leads to significant changes of the magnetic vortex texture in \textit{planar} disks. Namely, the intrinsic DMI of surface type favours the magnetization curling with the rotation direction being determined by the sign of the Dzyaloshinskii constant resulting in the increase or decrease of the vortex core size depending on the vortex chirality. 
In the following, we discuss the influence of the curvature-induced DMI on the properties of the magnetic vortex in a cap and compare the predicted effects to those induced by the intrinsic DMI.

\begin{figure}
	\includegraphics[width=\columnwidth]{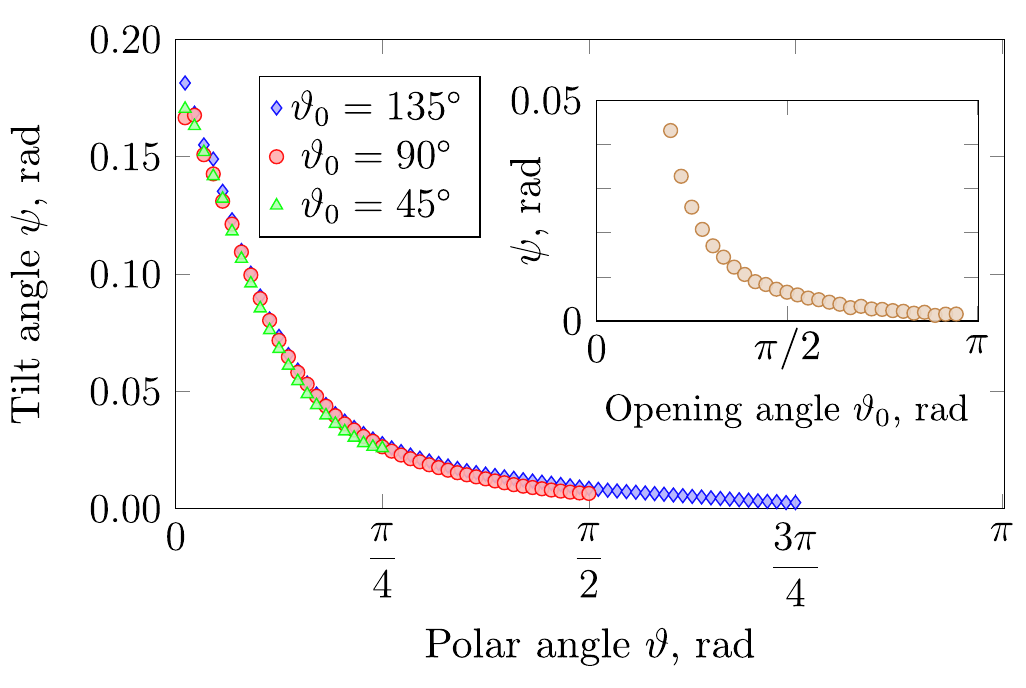}
	\caption{(Color online) \textbf{Tilt angle $\psi$ for a spherical cap.}
		 Dependence of the tilt angle $\psi$ on the polar angle $\vartheta$ for a spherical cap with inner radius $L=50$ nm, thickness $h = 5$ nm and different opening angles $\vartheta_0$. Inset shows how the tilt $\psi$ at the cap edge depends on cap opening angle $\vartheta_0$. The plots are obtained from micromagnetic simulations.}
	\label{fig:angularCorrectionVsTheta}
\end{figure}

It is instructive to start the analysis for spherical shells whose size allows to support strongly inhomogeneous magnetic textures instead of spontaneous formation of ferromagnetic domains~\cite{Sloika17}. For the case of a Heisenberg easy-surface ferromagnet shaped in a spherical shell geometry, the vortex state is realized only with the same polarity on both poles (the vortex polarity is defined with respect to the surface normal). A symmetry of the magnetic texture forces the equatorial magnetization to be strictly parallel to the equator of the sphere~\cite{Kravchuk12a}. This symmetry is absent for the case of a magnetically soft spherical cap, which can host the only one vortex. Therefore, it is to be expected that the edge magnetization $\vec{m}$ experiences tilt by an angle $\psi(\vartheta_0)$, which is dependent on the opening angle $\vartheta_0$ of the cap, see Fig.~\ref{fig:SphericalCap}. To determine this tilt, we carried out series of micromagnetic simulations using \textsf{magpar} code~\cite{MAGPAR, Scholz03a}. Technically, we performed energy minimization for spherical caps with the inner radius $L=50$ \unit{nm} and shell thickness $h=5$ \unit{nm} for different opening angles $\vartheta_0$. For simulations, material parameters similar to Permalloy (Py, \ch{Ni81Fe19}) were used: exchange constant $A = 10.5$ \unit{pJ/m} and  saturation magnetization $M_{\textsc{s}} = 795$ \unit{kA/m}. These parameters result in the exchange length $\ell\approx 5.14$ \unit{nm}. We utilize the spherical angle parametrization for the spherical curvilinear coordinates $(\vartheta,\chi, r)$ of the magnetization unit vector $\vec{m}$, namely, $m_1\equiv m_\vartheta = \sin\theta\cos\phi$, $m_2\equiv m_\chi = \sin\theta\sin\phi$, and $m_n\equiv m_r = \cos\theta$. For the vortex distribution in a thin spherical cap 
\begin{equation}
 \theta=\theta(\vartheta), \;\;\;\; \phi=\mathfrak{C} \left(\dfrac{\pi}{2} + \psi \right),
\end{equation}
where $\psi = \psi(\vartheta)$ represents the tilt angle and $\mathfrak{C}=\pm1$ determines the vortex circulation. Fig.~\ref{fig:angularCorrectionVsTheta} summarizes the results of micromagnetic simulations on the determination of the magnetization tilt at the edge of the cap for different opening angles $\vartheta_0$. The tilt angle $\psi$ has not zero value near the cap edge and increases in the vicinity of the vortex core. At the same time, an increase of the cap opening angle $\vartheta_0$ does not lead to significant changes of the tilt angle $\psi$ near the vortex core. On the other hand, increasing the opening angle results in the decrease of the tilt angle at the edge, see inset in Fig.~\ref{fig:angularCorrectionVsTheta}. This effect can be explained by the increase of the distance between the vortex core and cap edge  with the increase of the opening angle.

To describe the tilt angle in a spherical cap theoretically, it is convenient to use the following notation, which corresponds to the stereographic projection on a plane: 
\begin{equation} \label{eq:notation}
\rho =\frac{1}{R}\tan\frac{\vartheta}{2},\qquad \rho\in[0,1), \qquad R = \dfrac{1}{L} \tan\frac{\vartheta_0}{2}.
\end{equation}
Using this notation, one can write the Euler--Lagrange equations for the spherical geometry \cite{Kravchuk12a} in the following form:
\begin{subequations}\label{eq:EulerLagrange}
\begin{equation}\label{eq:EulerLagrange-theta}
	\theta'' + \frac{\theta'}{\rho} + \sin\theta \cos\theta \left(\mathcal{K} - \frac{1}{\rho^2} \right) - \mathcal{D} \frac{\sin^2\theta }{\rho} = 0,
\end{equation}
\begin{equation}\label{eq:EulerLagrange-psi}
	\psi'' + \frac{\psi'}{\rho}  + 2 \cot\theta\,  \theta' \psi' + 2 \cos\psi \frac{2 R}{1+R^2 \rho^2} \theta' =0,
\end{equation}
\end{subequations}
where prime symbols correspond to the derivative with respect to $\rho$. Effective anisotropy $\mathcal{K}$ and effective DMI $\mathcal{D}$ are functions of the coordinate $\rho$:
\begin{equation}
	\label{eq:InducedDMI&Aniso}
	\begin{split}
		&\mathcal{K}(\rho)=  \frac{R^2}{\beta^2(1+R^2\rho^2)^2} -(\psi^\prime)^2, \; \beta = \dfrac{\ell}{2\sqrt{L^2+2\ell^2}},\\
		&\mathcal{D}(\rho)= 4 \left(\frac{\rho R}{1+\rho^2 R^2} \sin\psi \right)^\prime,
\end{split}
\end{equation}
Eq.~\eqref{eq:EulerLagrange-theta} resembles a typical angular dependence of an axis-symmetric magnetic soliton in the easy-plane film with the coordinate-dependent anisotropy $\mathcal{K}$ and DMI $\mathcal{D}$. 

The key novelty of this manuscript is in the coupling of  Eq.~\eqref{eq:EulerLagrange-psi} with Eq.~\eqref{eq:EulerLagrange-theta} by the effective anisotropy and DMI. It is important to mention that the coordinate dependence of the effective DMI constant $\mathcal{D}$ does not lead to the driving on magnetic solitons. This is in contrast to the case of the skyrmion equation in curvilinear shells of more complex geometry~\cite{Pylypovskyi18a,Kravchuk18a}. The difference is due to the constant curvature of the geometry (spherical shell).

The curvature-induced DMI and anisotropy depend on the reduced polar angle $\rho$. In the limiting case of a disk ($\vartheta_0 \to 0$ keeping the product $\vartheta_0L = \text{const}$) the solution of the second equation in Eq.~\eqref{eq:EulerLagrange} is $\psi = 0$ (tilt angle disappears for planar disks). Consequently, $\mathcal{D}=0$, which means that the curvature-induced DMI disappears for planar disks.

\begin{figure}
 	\includegraphics[width=\columnwidth]{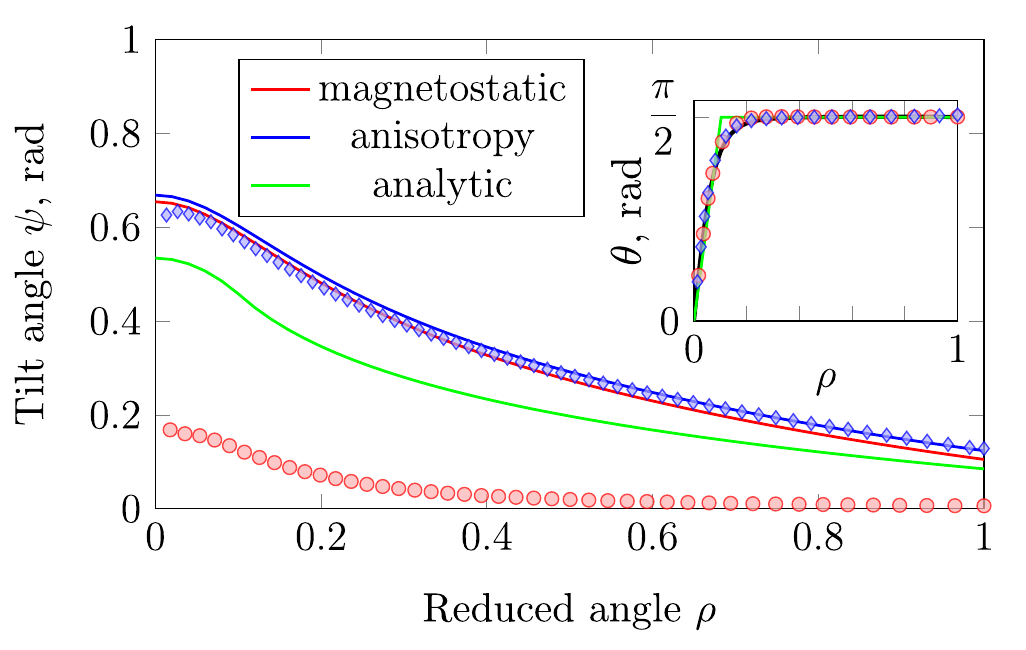}
 	\caption{(Color online) \textbf{Magnetic vortex parameters vs reduced angle $\rho$.} Comparison of the results of micromagnetic simulations (symbols) with the analytic and numerical solution (solid lines) of the Euler--Lagrange equations \eqref{eq:EulerLagrange}. Red circles correspond to the simulation results for the model with full scale magnetostatics. Blue diamonds correspond to the simulation results for the model with easy-surface anisotropy. Red and blue solid lines correspond to the numerical solution of Eq.~\eqref{eq:EulerLagrange} with the pinning parameter $d=0.47$ (model with full scale magnetostatics) and $d=0.52$ (model with easy-surface anisotropy), respectively. Green line corresponds to the analytic solution \eqref{eq:phi_solution}. Inset shows the out-of-surface component of magnetization $\theta$ for each of these models. The shown dependencies are calculated for the Permalloy spherical cap with inner radius $L=50$ nm, thickness $h=5$ nm and opening angle $\vartheta_0={\pi}/{2}$.}
 	\label{fig:vortexVsPho}
 \end{figure}

The equations~\eqref{eq:EulerLagrange} can be solved numerically with the boundary condition $\theta(0)=0$, $\theta(1)=\dfrac{\pi}{2}$, $\psi^\prime(0)=0$ and $\psi(1) + d \psi^\prime(1)=0$, where $d$ describes a pinning effect introduced by the surface energy at the cap edge. The parameter $d$ can be found from micromagnetic simulations. For Py spherical cap with inner radius $L=50$ nm, thickness $h=5$ nm and opening angle $\vartheta_0=\dfrac{\pi}{2}$, the pinning parameter $d \approx 0.47$. The numerical solution of Eq.~\eqref{eq:EulerLagrange} is shown in Fig.~\ref{fig:vortexVsPho} (red line in the main figure and black line in the inset). The obtained out-of-surface component $\theta(\rho)$ is in good agreement with the result of full scale micromagnetic simulations (red symbols). On the other hand, the numerical solution for the tilt angle $\psi$ differs significantly from the full scale micromagnetics. This can be explained by the fact that Eq.~\eqref{eq:EulerLagrange} are written for the Heisenberg easy-surface ferromagnet without nonlocal magnetostatic effects. In contrast, the full scale magnetostatic simulations account for not only surface but also volume charges that increase the total energy. To verify this assumption, we performed additional simulations of a Heisenberg easy-surface ferromagnet shaped as a spherical cap with the following parameters: exchange constant $A = 10.5$ \unit{pJ/m} and anisotropy constant $K = 398$ \unit{kJ/m^3}. These parameters result in the  magnetic length $\ell\approx 5.14$ \unit{nm} (the same value as for
the exchange length above). The tilt angle $\psi$ obtained from simulations of this model is displayed in Fig.~\ref{fig:vortexVsPho} by blue diamonds, which fits the numerical solution of Eq.~\eqref{eq:EulerLagrange} well. We observe that volume charges that are not taken into account in the later simulations of a cap with the easy-surface anisotropy lead to the decrease of the tilt angle $\psi$ while vortex core profile remains the same (blue diamonds in the inset of Fig.~\ref{fig:vortexVsPho}). Simulations of the model with the effective easy-surface anisotropy allow estimating the pinning parameter $d=0.52$. 
 
To analyse the tilt angle analytically, we use the following model for the angular parameter of the out-of-surface component of the magnetic vortex texture:
\begin{equation}\label{eq:theta_ansatz}
\theta =  
\begin{dcases}
\frac{\pi}{2}p \left(\frac{\rho}{\rho_c}-1\right) + \frac{\pi}{2}, & \rho < \rho_c\\
\frac{\pi}{2}, & \rho \ge \rho_c,
\end{dcases}
\end{equation}
where $p$ determines the vortex polarity and $\rho_c$ determines the size of the vortex core. In the case of a planar nanomagnet, the size of the vortex core can be determined as $2 \ell$. Consequently, for a spherical cap, we can expect $\rho_c \approx \tan \left({\ell}/{L}\right)$.
Substituting Eq.~\eqref{eq:theta_ansatz} into \eqref{eq:EulerLagrange-psi}, we find the solution for the case $\rho_c \ll 1$:
\begin{equation}\label{eq:phi_solution}
\psi =  
\begin{dcases} 
\psi^< = \phi_1 -\phi_0 \rho^2, & \rho < \rho_c\\
\psi^> = \phi_2 \ln \rho + \phi_3, & \rho \ge \rho_c,
\end{dcases}
\end{equation}
where $\phi_0$ can be found from Eq.~\eqref{eq:EulerLagrange-psi} and constants $\phi_1, \phi_2, \phi_3$ can be found from the condition $$\left[\dfrac{\psi\prime}{\psi}\right]_{\rho_c}=0, \;\;\;\; \psi(1) + d  \psi^\prime(1) = 0.$$ 
Finally, constants in Eq.~\eqref{eq:phi_solution} read:
\begin{equation} \label{eq:phi_solution_consts}
\begin{aligned}
\phi_0 &=\frac{\pi}{4} \frac{1}{\rho_c} p R,&  \phi_1 &=  \dfrac{\pi}{4} p R \rho_c \left(1+2d -2 \ln \rho_c \right)\\
\phi_2 &= -\frac{\pi}{2} p R \rho_c ,& \phi_3 &=  \frac{\pi}{2} p R \rho_c d.
\end{aligned}	
\end{equation}
The analytical solution, Eq.~\eqref{eq:phi_solution} (green line in Fig.~\ref{fig:vortexVsPho}), slightly underestimates the tilt angle $\psi$ compared to the numerical solution of Eq.~\eqref{eq:EulerLagrange} (blue line in Fig.~\ref{fig:vortexVsPho}) and simulation results (blue diamonds in Fig.~\ref{fig:vortexVsPho}) for the model with easy-surface anisotropy. The observed discrepancy is due to a rather crude model used for the description of the out-of-surface angular component of magnetization.


\section{Out-of-surface component of magnetization}

\label{sec:out-of-surface}
	
\citeauthor{Butenko09} analysed the impact of the \textit{intrinsic} DMI of volume type on vortices in planar disks~\cite{Butenko09}. The Euler--Lagrange equation for the out-of-surface magnetization obtained for the case of planar disks with the volume type DMI~\cite{Butenko09} has the same form as the Euler--Lagrange equation \eqref{eq:EulerLagrange} for $\theta$ for the case of spherical caps. Thus, it is insightful to compare the influence of the curvature-induced DMI on the out-of-surface magnetization component of vortices in \textit{achiral} spherical caps with the case of planar disks of a ferromagnet with an intrinsic DMI.

\begin{figure}
	\includegraphics[width=\columnwidth]{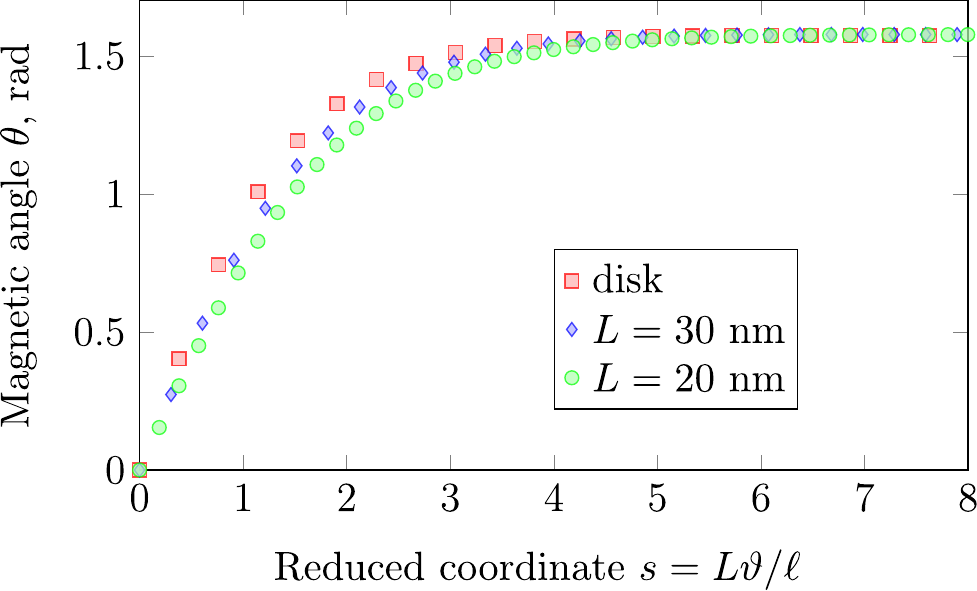}
	\caption{(Color online) \textbf{Vortex core profile for different curvature radii. } Symbols correspond to the magnetic angle $\theta$ (describes the out-of-surface magnetization component) obtained from simulations for different curvature radii: green circles correspond to a spherical cap with the inner radius $L=20$~nm, thickness $h=5$~nm and opening angle $\vartheta_0 = 175^{\circ}$; blue diamonds correspond to a spherical cap with the inner radius $L=30$~nm, thickness $h=5$~nm and opening angle $\vartheta_0 = 115^{\circ}$; red squares correspond to planar disk with radius $L=100$~nm and thickness $h=5$~nm.}
	\label{fig:ThetaVsVartheta}
\end{figure}

Taking into account that the Dzyaloshinskii constant $\mathcal{D}$ is determined by $L$, the vortex core profile should be dependent on the radius of the cap. This can be verified by comparing numerical solutions of Eq.~\eqref{eq:EulerLagrange} for the cases $\mathcal{D}\ne0$ (spherical cap) and $\mathcal{D}=0$ (disk). We performed micromagnetic simulations of the ground state for spherical caps with the thickness $h=5$~nm and different inner radii and opening angles. To compare the size of the vortex core for different caps, we rely on the reduced coordinate along the cap meridian $s = L \vartheta/\ell$, see Fig.~\ref{fig:ThetaVsVartheta}. We observe that the vortex core is slightly larger for the spherical cap with inner radius $L=20$~nm and opening angle $\vartheta_0 = 175^\circ$ (blue diamonds in Fig.~\ref{fig:ThetaVsVartheta}) than in the case of the cap with $L = 30$~nm and $\vartheta_0 = 115^\circ$ (green circles in Fig.~\ref{fig:ThetaVsVartheta}). For comparison, the vortex core profile for a planar disk is displayed in Fig.~\ref{fig:ThetaVsVartheta} (red squares). 

\section{Conclusion}
\label{sec:conclusion}

In conclusion, geometric curvature of a spherical cap leads to the appearance of the effective coordinate-dependent anisotropy and DMI even for the case of intrinsically isotropic and achiral ferromagnets. Here, we derive and analyze the tilt of the azimuthal angle of magnetization for the vortex texture for the spherical cap geometry. The presence of the curvature-induced DMI significantly modifies the properties of magnetic vortices. Namely, we observe a tilt of the magnetization at the edge of the cap as well as a slight increase of the size of the magnetic vortex core. The obtained analytical results are verified using micromagnetic simulations of the models with full scale magnetostatics and easy-surface anisotropy. 

\begin{acknowledgments}
The simulation results were obtained using the computing cluster of Taras Shevchenko National University of Kyiv \cite{unicc}. This work is financed in part via the German Research Foundation (DFG) under Grants No. MA 5144/14-1, MA 5144/22-1, MA 5144/24-1, MC 9/22–1. V.P.K. acknowledges the Alexander von Humboldt Foundation for the support and Leibniz IFW Dresden for kind hospitality. M.I.S. and D.D.S. acknowledge HZDR, where a part of work was performed, for kind hospitality. M.I.S. acknowledges DAAD (Leonhard-Euler-Programm) for financial support. O.V.P. acknowledges a support within the program ``Fuzzy continuous quantum measurements'' of NAS of Ukraine (KPKVK 6541040).
\end{acknowledgments}

%
%
%
%

\end{document}